\documentclass[12pt]{article}
\usepackage{graphicx}
\usepackage{amsmath,amssymb} 
\usepackage{xspace}
\usepackage{multicol}
\usepackage{multirow}%
\usepackage{hepparticles}
\usepackage{hyperref}
\providecommand{\doi}{\texttt{doi:}\begingroup \urlstyle{tt}\Url}
\providecommand{\DOI}[1]{\href{http://dx.doi.org/#1}{\doi{#1}}}
\newcommand\pubnumber{CMS CR-2023/296}
\newcommand\pubdate{\today}

\def\institute{Department of Physics\\
National Taiwan university}
\def\authemail{\footnote{Contact: ting-hsiang.hsu@cern.ch}}

\def\Title#1{\begin{center} {\Large #1 } \end{center}}
\def\Author#1{\begin{center}{ \sc #1} \end{center}}
\def\Address#1{\begin{center}{ \it #1} \end{center}}

\newcommand\pubblock{\rightline{\begin{tabular}{l} \pubnumber\\
         \pubdate  \end{tabular}}}
\newenvironment{Abstract}{\begin{quotation}  }{\end{quotation}}
\newenvironment{Presented}{\begin{quotation} \begin{center} 
             PRESENTED AT\end{center}\bigskip 
      \begin{center}\begin{large}}{\end{large}\end{center} \end{quotation}}

\usepackage{hepparticles}
\def\beq{\begin{equation}}
\def\eeq#1{\label{#1}\end{equation}}
\def\eeqn{\end{equation}}

\def\beqa{\begin{eqnarray}}
\def\eeqa#1{\label{#1}\end{eqnarray}}
\def\eeqan{\end{eqnarray}}

\let\bar=\overbar

\def\Dslash{\not{\hbox{\kern-4pt $D$}}}
\def\dslash{\not{\hbox{\kern-2pt $\del$}}}

\def\msb{{\bar{\ssstyle M \kern -1pt S}}}

\newcommand{\Xspace}{\xspace}

\providecommand{\PGmpm}{\ensuremath{\mu^\pm}\xspace}

\newcommand{\pt}{\ensuremath{p_{\mathrm{T}}}\xspace}
\newcommand{\PA}{\HepParticle{A}{}{}\Xspace}
\newcommand{\PH}{\HepParticle{H}{}{}\Xspace}
\newcommand{\PW}{\HepParticle{W}{}{}\Xspace}
\newcommand{\Ph}{\HepParticle{h}{}{}\Xspace}
\newcommand{\Pp}{\HepParticle{p}{}{}\Xspace}

\newcommand{\PQt}{{\HepParticle{t}{}{}}\Xspace}
\newcommand{\PQc}{{\HepParticle{c}{}{}}\Xspace}
\newcommand{\PQb}{{\HepParticle{b}{}{}}\Xspace}
\newcommand{\PQu}{{\HepParticle{u}{}{}}\Xspace}
\newcommand{\PQq}{{\HepParticle{q}{}{}}\Xspace}
\newcommand{\PAQt}{{\HepParticle{\bar{t}}{}{}}\Xspace}
\newcommand{\PAQc}{{\HepParticle{\bar{c}}{}{}}\Xspace}
\newcommand{\PAQu}{{\HepParticle{\bar{u}}{}{}}\Xspace}
\newcommand{\PAQq}{{\HepParticle{\bar{q}}{}{}}\Xspace}
\newcommand{\Pe}{{\HepParticle{{e}}{}{}}\Xspace}
\newcommand{\Pepm}{\ensuremath{\Pe^\pm}\xspace}
\newcommand{\ttc}{\ensuremath{\PQt{}\PQt{}\PAQc}\xspace}
\newcommand{\ttu}{\ensuremath{\PQt{}\PQt{}\PAQu}\xspace}
\newcommand{\ttq}{\ensuremath{\PQt{}\PQt{}\PAQq}\xspace}

\newcommand{\ttbar}{{\PQt{}\PAQt}\xspace} 
\newcommand{\ttW}{\ensuremath{\ttbar\PW}\xspace}

\newcommand{\ee}{\ensuremath{\Pepm\Pepm}\xspace}
\newcommand{\mumu}{\ensuremath{\PGmpm\PGmpm}\xspace}
\newcommand{\emu}{\ensuremath{\Pepm\PGmpm}\xspace}
\newcommand{\pp}{\ensuremath{\Pp\Pp}\xspace}

\newcommand{\abs}[1]{\ensuremath{\lvert #1 \rvert}}
\newcommand{\ptmiss}{\ensuremath{\pt^\text{miss}}\xspace}
\newcommand{\ptvecmiss}{\ensuremath{{\vec p}_{\mathrm{T}}^{\kern1pt\text{miss}}}\xspace}
\newcommand{\HT}{\ensuremath{H_{\mathrm{T}}}\xspace}
\newcommand{\GeV}{\ensuremath{\,\text{Ge\hspace{-.08em}V}}\xspace}

\newcommand{\TeV}{\ensuremath{\,\text{Te\hspace{-.08em}V}}\xspace}

\begin{document}
\begin{titlepage}
\pubblock

\vfill
\Title{Search for new Higgs bosons in same sign top-quark pair+jets final state}
\vfill
\Author{ Ting-Hsiang Hsu\authemail on behalf of CMS Collaboration}
\Address{\institute}
\vfill
\begin{Abstract}
A search is presented for new Higgs bosons, targeting proton-proton ($\pp$) collision events with a same-sign top quark pair associated with an extra jet via the processes $\pp\rightarrow \PQt\PH/\PA \rightarrow \ttc$ and $\pp\rightarrow \PQt\PH/\PA \rightarrow \ttu$, where $\PH$ and $\PA$ represent exotic scalar and pseudoscalar bosons, respectively. The study is based on data collected at a center-of-mass energy of 13 $\TeV$ with the CMS detector at LHC Run 2, corresponding to an integrated luminosity of 138 $\text{fb}^{-1}$. The analysis is made based on generalized two-Higgs-doublet model (g2HDM). It targets the new Higgs masses ranging from 200 $\GeV$ to 1$\TeV$ and extra Yukawa couplings, $\rho_{\PQt\PQu}$ and $\rho_{\PQt\PQc}$, from 0.1 to 1.0. Two scenarios are studied, in which only one of $\PH$ and $\PA$ exists or in which they coexist and interfere. No significant excess above standard model predictions is observed.
\end{Abstract}
\vfill
\begin{Presented}
$16^\mathrm{th}$ International Workshop on Top Quark Physics\\
(Top2023), 24--29 September, 2023
\end{Presented}
\vfill
\end{titlepage}
\def\thefootnote{\fnsymbol{footnote}}
\setcounter{footnote}{0}

\section{Introduction}
\begin{figure}[!h!tbp]
\centering
\includegraphics[height=1.5in]{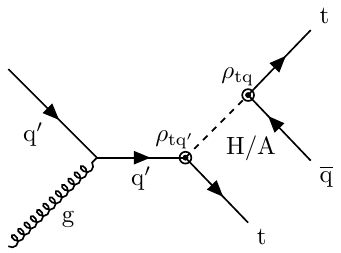}
\caption{Representative Feynman diagram for tt$\bar{q}$ (q = u, c) production through a new scalar(H) or pseudoscalar(A) Higgs boson. In this analysis, events with $q=q^\prime$ are considered. Figure is taken from Ref~\cite{CMS:2023xpx}.}
\label{fig:feynman_diag}
\end{figure}

The ATLAS~\cite{ATLAS:2008xda} and CMS~\cite{CMS:2008xjf} experiments discovered a scalar boson, $\Ph_{125}$, with a mass of 125$\GeV$, using proton-proton ($\pp$) collision data collected during the CERN LHC Run 1 (2011-2012)~\cite{Aad:2012tfa,Chatrchyan:2012ufa}. The scalar boson has the properties that are so far consistent with the Higgs boson predicted by the standard model (SM)~\cite{Khachatryan:2016vau}. Natural questions then emerge about the existence of additional SU(2) doublets. With introducing a second doublet, two-Higgs-doublet model (2HDM) predicts five physical scalar bosons: CP-even neutral scalar bosons $\Ph$ and $\PH$ with $m_{\Ph} < m_{\PH}$, CP-odd neutral pseudoscalar boson $\PA$, and two charged scalar bosons $\PH^\pm$~\cite{Branco:2011iw}. These scalar bosons with sub$\TeV$ mass scales can be searched with the existing LHC data~\cite{Kohda:2017fkn}, but their signatures may be suppressed by fermion mass-mixing hierarchy~\cite{Hou:1991un} and alignment mechanisms, where the mixing angle $\gamma$ between the scalar bosons $\Ph$ and $\PH$ has a value for which $\cos{\gamma}\approx 0$ ~\cite{Hou:2017hiw}. In the alignment limit~\cite{Gunion:2002zf}, the lighter scalar neutral boson $\Ph$ and SM $\Ph_{125}$ boson become the same and its flavor changing neutral currents (FCNC) are suppressed. This alignment scenario may be achieved if extra Higgs quartic couplings are $\mathcal{O}(1)$ and no $\mathbb{Z}_2$ symmetry requirement is imposed~\cite{Hou:2017hiw}, allowing FCNC involving H and A bosons. This scenario without $\mathbb{Z}_2$ symmetry can be studied in the framework of the generalized 2HDM (g2HDM)~\cite{Lee:1973iz, Hou:1991un, Branco:2011iw}. In this model, new Yukawa couplings emerge and some of them, such as $\rho_{\PQt\PQt}, \rho_{\PQc\PQc}$, may still be assumed to have large values. These couplings combined with $\mathcal{O}(1)$ Higgs quartic couplings may explain the electroweak baryogenesis~\cite{Fuyuto:2017ewj,Fuyuto:2019svr}. The new top quark Yukawa coupling $\rho_{\PQt\PQc}$ with $\rho_{\tau\mu}$ having a similar strength may also explain a possible muon g-2 anomaly~\cite{Muong-2:2023cdq} while being compatible with the observed data depending on the $\PH^\pm$ mass.\\
 In this paper, we present a search for the existence of the real part of these couplings, $\rho_{\PQt\PQu}$ and $\rho_{\PQt\PQc}$ through the $\pp\rightarrow \PQt\PH/\PA\rightarrow \ttq(\PQq=\PQu,\PQc)$ and its charge conjugate processes, considering only one coupling at a time~\cite{CMS:2023xpx}. A representative Feynman diagram for the $\ttq$ process is displayed in Fig\ref{fig:feynman_diag}. The analysis is based on the $\pp$ collision data collected at a center-of-mass energy of 13 $\TeV$  with the CMS detector at the LHC Run 2  (2016-2018), corresponding to an integrated luminosity of 138 $\text{fb}^{-1}$.\\

\section{Event reconstruction and selection}
Data is selected online with double lepton and single lepton triggers to maximize the signal efficiency. A multivariant discriminant~\cite{CMS:2020mpn} and a cut based technique are used for electron and muon identification, respectively. For improving the measurement of the lepton charge, further criteria is imposed~\cite{Khachatryan:2015hwa}. An additional multivariant discriminant is introduced for both electrons and muons in order to improve the separation of prompt leptons from nonprompt leptons that originate from decays of hadrons in the jets, hadrons misidentified as leptons, or photon conversions~\cite{CMS:2022tkv}. ``Tight" electrons(muons) are required to have $\pt > 20$ $\GeV$ and $\abs{\eta} < 2.5(2.4)$ while ``loose" ones $\pt > 10$ $\GeV$ with the same $\abs{\eta}$ condition. The electrons falling in the gap between ECAL barrel and endcap region ($1.442 < \abs{\eta} < 1.554$) are vetoed. Jets are required to have $\pt > 30$ $\GeV$, $\abs{\eta} < 2.4$ and separated from leptons by at least $\Delta R(j,\ell)=0.4$, where $\Delta R(j,\ell) = \sqrt{(\Delta\eta(j,\ell))^2 + (\Delta\phi(j,\ell))^2}$. To identify different jet flavors, variables CvsB and CvsL~\cite{CMS:2021scf,CMS-DP-2023-006}, which indicates the ratio of the probability of the jets to be from $\PQc$ jets over that from $\PQb$ jets or from light jets and gluons, respectively, are used. These variables are provided by the \textsc{DeepJet} algorithm~\cite{Bols:2020bkb}, a neural network utilizing
global variables, charged and neutral particle, and secondary
vertex features in a jet to perform flavor tagging.\\
The events are required to have exactly two same-sign charge ``tight" leptons, in three channels: $\ee, \emu, \mumu$.
The ones with the third lepton passing ``loose" criteria are dropped. They are also required to have at least three jets
and $\ptmiss > 35 \GeV$. The two leptons are required to have $\Delta R(\ell,\ell) > 0.3$ and invariant mass $m(\ell,\ell)>20\GeV$. For the $\ee$ channel, events with $60 \GeV < m(\ell,\ell) <  120\GeV$ are vetoed to suppress Drell-Yan backgrounds.\\
The nonprompt background is estimated from control regions utilizing data through the fake-factor method~\cite{CMS:2018fdh}. A similar method is applied to estimate charge misidentified background in the $\ee$ channel. Other backgrounds are estimated via Monte-Carlo simulations.
\section{Signal extraction}
A boosted decision tree (BDT) discriminant~\cite{Friedman:2001wbq} is used to separate the signal from the background. Utilizing half of the Monte-Carlo samples, the training is performed using input variables listed in table~\ref{tab:bdt_inputs}. For nonprompt background, semileptonic $\ttbar$ Monte-Carlo sample is used as a proxy in the BDT training. \\

\begin{table*}[!htpb] 
\centering 
\begin{tabular}{lll} 
  \multicolumn{2}{c}{Input variables of the BDT} & \\
\hline
CvsL$(j_a)$ & $a=1,2,3$ & Charm- vs light-quark jet flavor\\ & & identification variable\\ 
CvsB$(j_a)$ & $a=1,2,3$ & Charm- vs bottom-quark jet flavor\\ & &identification variable\\
$\Delta R(j_a,j_b)$ & $1\leq a<b\leq3$ & Angular separation between jets\\  
$m(j_a,j_b)$ & $1\leq a<b\leq3$ & Invariant mass of jet pairs\\
$\Delta R(j_a,l_b)$ & $a=1,2,3$; $b=1,2$ & Angular separation between jet and lepton\\ 
$m(j_a,l_b)$ & $a=1,2,3$; $b=1,2$ & Invariant mass of jet-lepton pairs\\
$\pt(\ell_a)$ & $a=1,2$ & Transverse momentum of leptons\\ 
$m(\ell_1,\ell_2,j_a)$ & $a=1,2,3$ & Invariant mass of the two leptons\\ & & plus the highest $\pt$ jet\\
$m(\ell_1,\ell_2)$ & & Invariant mass of the two leptons\\ 
\HT & & Scalar $\pt$ sum of the jets\\ 
\ptmiss & & Missing transverse momentum\\
\end{tabular} 
\caption{Input variables of the BDT. Jets and leptons are ordered by \pt. } 
\label{tab:bdt_inputs} 
\end{table*}
To maximize the sensitivity, independent BDTs are trained in four eras of data-taking and also for different H/A mass assumptions: $m_{\PA/\PH}$ from 200 to 1000 $\GeV$ in a scenario where either $\PH$ or $\PA$ exists, and $m_{\PA}$ from 250 to 1000 $\GeV$ where H and A coexist and interfere ($m_{\PA}-m_{\PH}=50$ $\GeV$ is assumed). Also, BDTs are trained separately for $\rho_{\PQt\PQu}$ and $\rho_{\PQt\PQc}$ cases, for which coupling value is assumed to be 0.4 in each case. Signals with different coupling values are obtained through scaling the ones with the coupling value $0.4$.\\
The signal strength parameter, $\hat{\mu}$, is obtained by a simultaneous maximum likelihood fit~\cite{CMS-NOTE-2011-005} performed on three decay channels in four eras for each mass and coupling assumption. BDT score $>-0.6$ cut is applied in order to reduce the impact on the fit of background-dominated regions and also help to improve the stability of the fit and corresponding fit uncertainties.
\section{Systematic uncertainties}
Systematic uncertainties arise from various sources, such as detector effects, mis-modeling and theoretical uncertainties. These effects modify the event yield or the shape of measured distribution. The systematic uncertainties are modeled as nuisance parameters when performing maximum likelihood estimation to determine the best fit signal strength $\hat{\mu}$~\cite{CMS-NOTE-2011-005}.\\
The systematic uncertainties can be categorized in two main groups: experimental uncertainties and theoretical uncertainties. In this study, the experimental uncertainties are those related to the integrated luminosity, pileup, L1 trigger inefficiency, nonprompt-lepton and charge misidentified background estimation, jet energy scale and resolution, unclustered energy scale, lepton identification, muon momentum scale, jet flavor identification and trigger efficiencies. The theoretical uncertainties are those related to matrix-element renomalization and factorization scales, parton shower scales that control the initial- and final-state radiation, parton distribution functions, and, for backgrounds estimated from Monte-Carlo samples, uncertainties of cross sections. The dominant uncertainties come from nonprompt-lepton background estimation, $\ttW$ cross
section, jet flavor identification and statistics.
\section{Results}
\begin{figure}[!h!tbp]
\centering
\includegraphics[height=0.45\textwidth]{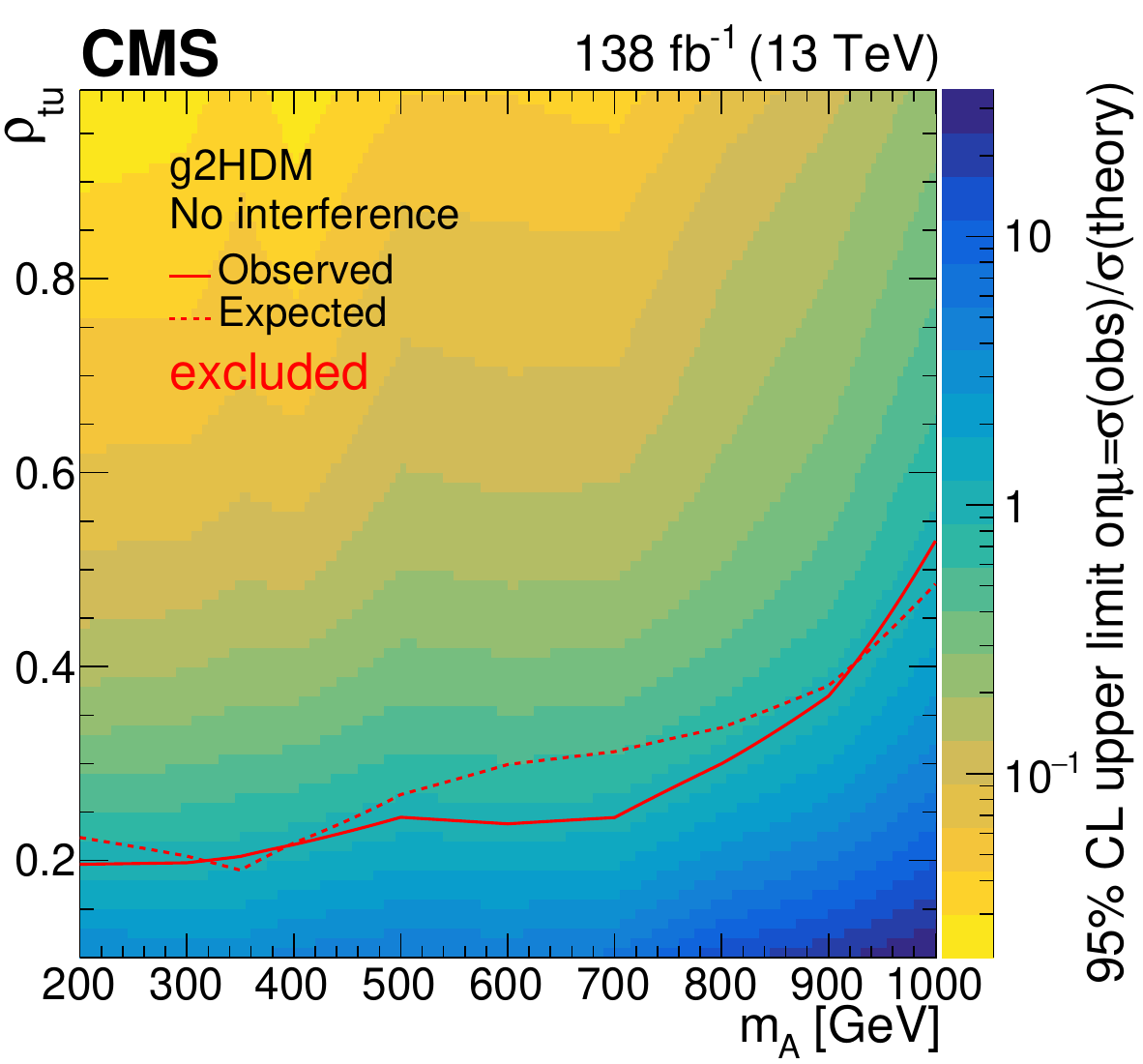}
\includegraphics[height=0.45\textwidth]{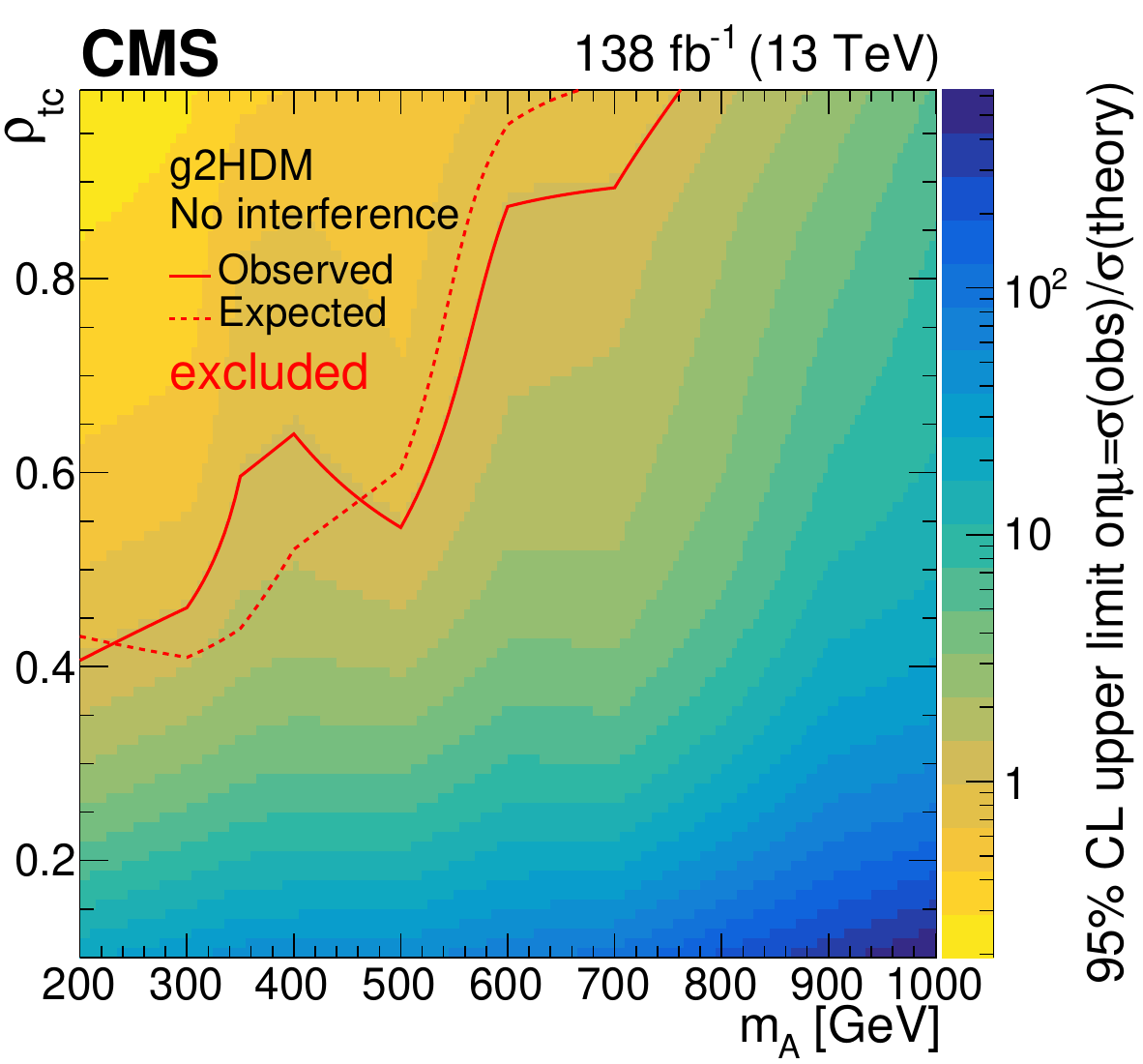}
\caption{Expected and observed upper limit as a function of $m_{\PA}$ from 200$\GeV$ to 1$\TeV$ for extra Yukawa coupling $\rho_{\PQt\PQu}$(left) and $\rho_{\PQt\PQc}$(right) from 0.1 to 1.0 in the scenario where only $\PH$ or $\PA$ exists. Figure is taken from Ref~\cite{CMS:2023xpx}.}
\label{fig:2Dlimit_noninterference}
\end{figure}

\begin{figure}[!h!tbp]
\centering
\includegraphics[height=0.45\textwidth]{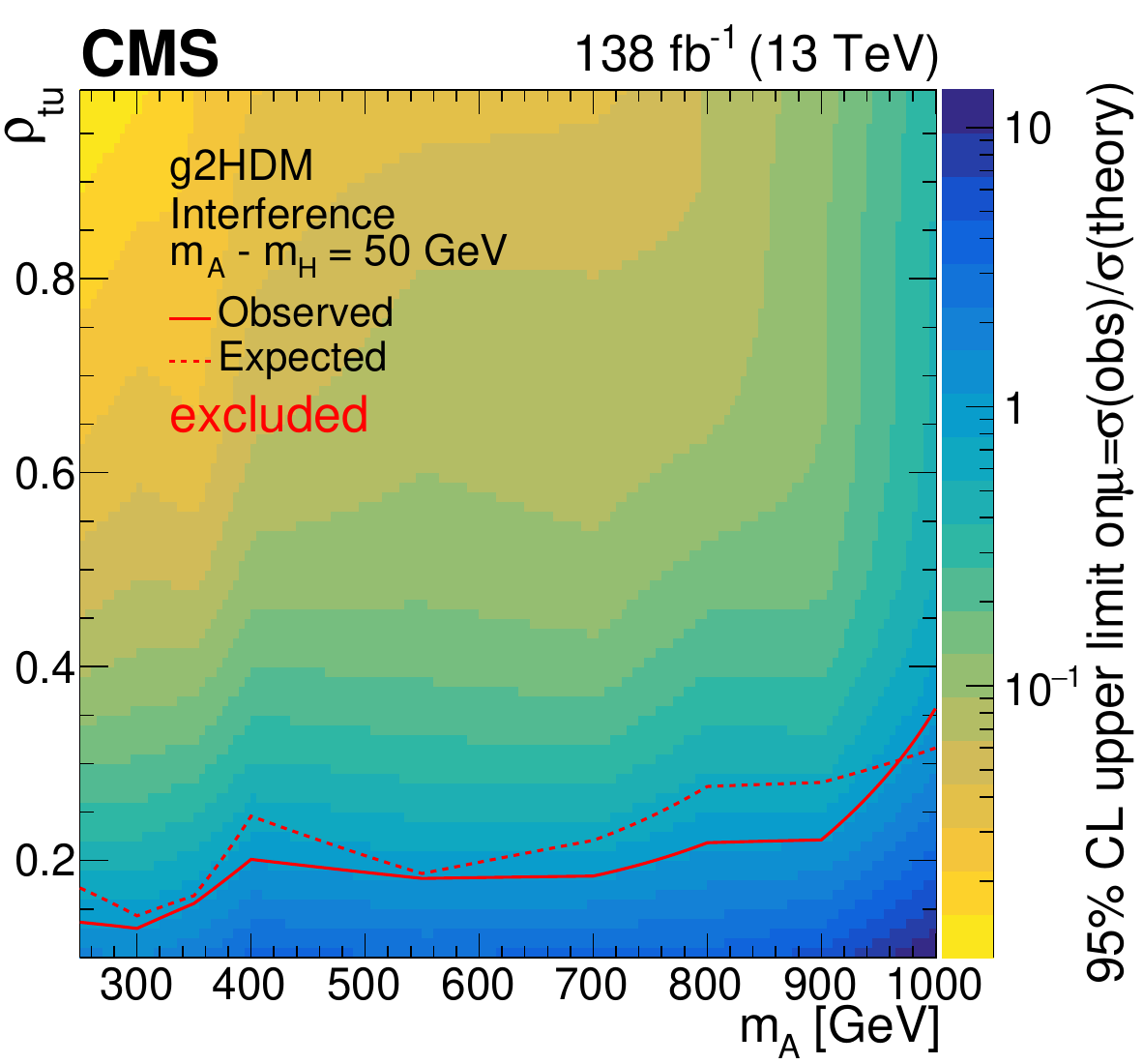}
\includegraphics[height=0.45\textwidth]{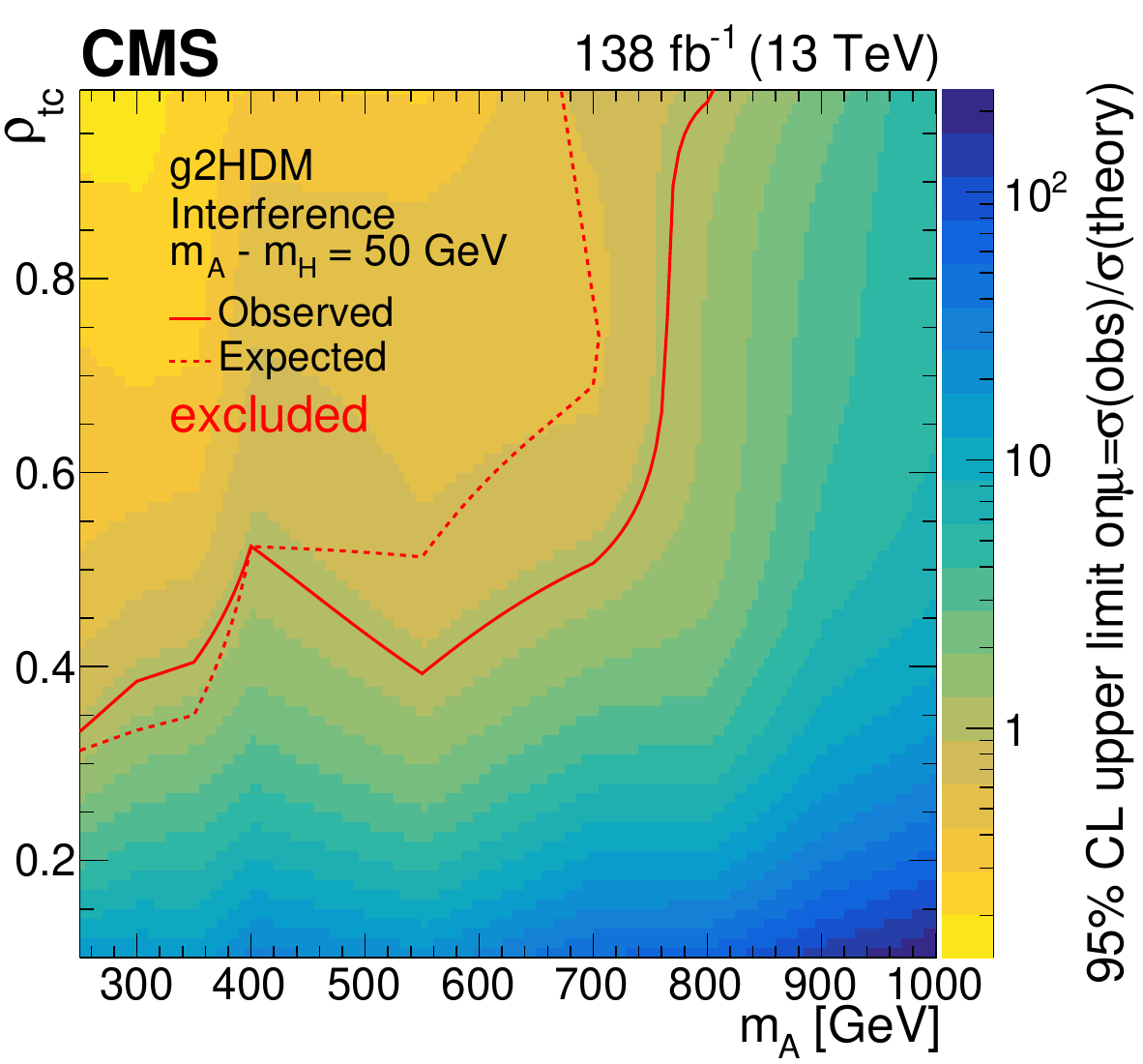}
\caption{Expected and observed upper limit as a function of $m_{\PA}$ from 200$\GeV$ to 1$\TeV$ and extra Yukawa coupling $\rho_{\PQt\PQu}$(left) and $\rho_{\PQt\PQc}$(right) from 0.1 to 1.0 in the scenario where $\PH$ and $\PA$ coexist and interfere with mass difference of 50$\GeV$. Figure is taken from Ref~\cite{CMS:2023xpx}.}
\label{fig:2Dlimit_interference}
\end{figure}

The results of the study are interpreted in terms of upper limit of signal strength, $\hat{\mu}$, in the context of g2HDM. The upper limits are calculated at 95\% confidence level (CL) and with an asymptotic approximation~\cite{bib:CLS3}. It is observed that in the scenario where only $\PH$ or $\PA$ exist, $\PH$ and $\PA$ can be used interchangeably since they have the same BDT distribution and also the same signal cross sections. \\
Figure~\ref{fig:2Dlimit_noninterference} and ~\ref{fig:2Dlimit_interference} show the observed 95\% CL upper limit as a function of $m_A$ and extra Yukawa coupling $\rho_{\PQt\PQu}$, $\rho_{\PQt\PQc}$ for the scenario with(without) $\PA$--$\PH$ interference. No significant excess above the standard model predictions is observed.

\bibliography{eprint}{}

\end{document}